\documentclass[10pt,journal,compsoc]{IEEEtran}
\ifCLASSOPTIONcompsoc
  \usepackage[nocompress]{cite}
\else
  \usepackage{cite}
\fi

\ifCLASSINFOpdf
   \usepackage[pdftex]{graphicx}
\else
\fi

\usepackage{amsmath}
\usepackage{algorithm}
\usepackage{algpseudocode}
\algnewcommand\algorithmicinput{\textbf{Input:}}
\algnewcommand\algorithmicoutput{\textbf{Output:}}
\algnewcommand\algorithmicbl{\textbf{\# of blocks:}}
\algnewcommand\algorithmicth{\textbf{\# of threads / block:}}
\algnewcommand\Input{\item[\algorithmicinput]}
\algnewcommand\Output{\item[\algorithmicoutput]}
\algnewcommand\Block{\item[\algorithmicbl]}
\algnewcommand\Thread{\item[\algorithmicth]}
\algrenewcommand\algorithmicindent{1.0em}

\usepackage{pifont}
\usepackage{tabularx}
\usepackage{array,booktabs}
\usepackage{multirow}
\usepackage{slashbox}
\usepackage{dblfloatfix}
\newcolumntype{C}{>{\centering\arraybackslash}X} 
\usepackage{lipsum}

\usepackage{soul}
\usepackage{array}
\usepackage{color}
\usepackage{nicefrac}

\newcolumntype{P}[1]{>{\centering\arraybackslash}p{#1}}

\newcommand\independent{\protect\mathpalette{\protect\independenT}{\perp}}
\def\independenT#1#2{\mathrel{\rlap{$#1#2$}\mkern2mu{#1#2}}}

\newcommand{\mStoredSepSet}{SepSet}
\newcommand{\mL}{\ell}
\newcommand{\mSetVar}{\mathcal{V}}
\newcommand{\mCIAsser}{\mathcal{I}}
\newcommand{\mCITest}{I}
\newcommand{\mTh}{\tau}
\newcommand{\mSh}{A'_{sh}}

\newcommand{\mEB}{\beta}
\newcommand{\mET}{\gamma}
\newcommand{\mSB}{\delta}
\newcommand{\mST}{\theta}
\newcommand{\mCombFunc}{Comb}

\newcommand{\mRefAlg}[1]{Algorithm~\ref{#1}}
\newcommand{\mRefFig}[1]{Fig.~\ref{#1}}
\newcommand{\mRefEq}[1]{Equation~\ref{#1}} 

\newcommand{\mRefSec}[1]{Section~\ref{#1}}

\usepackage{hyperref}

\begin{document}

\title{cuPC: CUDA-based Parallel PC Algorithm for Causal Structure Learning on GPU}
%
%
%
%

\author{Behrooz~Zarebavani,~
        Foad~Jafarinejad,~
        Matin~Hashemi,~
        and~Saber~Salehkaleybar
        
{\color{blue} 
	\begin{flushleft}
		\footnotesize 
		This article is published. Please cite as B. Zarebavani, F. Jafarinejad,  M. Hashemi, S. Salehkaleybar, ``cuPC: CUDA-based Parallel PC Algorithm for Causal Structure Learning on GPU," IEEE Transactions on Parallel and Distributed Systems (TPDS), 2019. doi: 10.1109/TPDS.2019.2939126
\end{flushleft} }        
        
\thanks{
The authors are with Learning and Intelligent Systems Laboratory, Department of Electrical Engineering, Sharif University of Technology, Tehran, Iran. Webpage: http://lis.ee.sharif.edu/ 
E-mails: behrooz.zare@ee.sharif.edu, fzj5053@psu.edu, matin@sharif.edu (corresponding author),  saleh@sharif.edu. \protect \\
doi: 10.1109/TPDS.2019.2939126
}
}

%
%

\markboth{IEEE Transactions on Parallel and Distributed Systems}%
{IEEE Transactions on Parallel and Distributed Systems}
\IEEEtitleabstractindextext{%
\begin{abstract}
The main goal in many fields in the empirical sciences is to discover causal relationships among a set of variables from observational data. PC algorithm is one of the promising solutions to learn underlying causal structure by performing a number of conditional independence tests. In this paper, we propose a novel GPU-based parallel algorithm, called cuPC, to execute an order-independent version of PC. The proposed solution has two variants, cuPC-E and cuPC-S, which parallelize PC in two different ways for multivariate normal distribution. Experimental results show the scalability of the proposed algorithms with respect to the number of variables, the number of samples, and different graph densities. For instance, in one of the most challenging datasets, the runtime is reduced from more than $11$ hours to about $4$ seconds. On average, cuPC-E and cuPC-S achieve $500$ X and $1300$ X speedup, respectively, compared to serial implementation on CPU. The source code of cuPC is available online  \cite{sourcecode}. 
\end{abstract}

}

\maketitle

\IEEEdisplaynontitleabstractindextext

%
\IEEEpeerreviewmaketitle

\section{Introduction}
\label{sec:intro}

\IEEEPARstart{L}{earning} causal structures is one of the main problems in empirical sciences. For instance, we need to understand the impact of a medical treatment on a disease or recover causal relations between genes in gene regulatory networks (GRN)~\cite{friedman2000using}. By discovering such causal relations, one will be able to predict the impact of different actions. Causal relations can be inferred by controlled randomized experiments. However, in many cases, it is not possible to perform the required experiments due to technical or ethical reasons. In such cases, causal relations need to be learned merely from observational data~\cite{pearl2003causality,Spirtes2000}. 

Causal Bayesian network is one of the models which has been widely considered to explain the data-generating mechanism. In this model, causal relations among variables are represented by a directed acyclic graph (DAG) where there is a direct edge from variable $V_i$  to variable $V_j$ if $V_i$ is a direct cause of $V_j$. The task of causal structure learning is to learn all DAGs that are compatible with the observed data. Under some assumptions \cite{Spirtes2000}, the underlying true causal structure is in the set of recovered DAGs if the number of observed data samples goes to infinity. 
Two common approaches for learning causal structures are score-based and constraint-based approaches. In the score-based approach, in order to find a set of DAGs that best explains dependency relations among the variables, a score function is evaluated, which might become an NP-hard problem~\cite{chickering1994learning}. 

In the constraint-based approach, such DAGs are found by performing a number of conditional independence (CI) tests. Sprites and Glymour \cite{Spirtes2000} proposed a promising solution, called PC algorithm. For ground-truth graphs with bounded degrees, PC algorithm does not require to perform high-order conditional independence tests, and thus, runs in polynomial time. PC algorithm has become a common tool for causal explorations and is available in different graphical model learning packages such as pcalg~\cite{kalisch2012causal}, bnlearn~\cite{scutari2009learning}, and TETRAD~\cite{tetrad}. Moreover, it has been widely applied in different applications such as learning the causal structure of GRNs from gene expression data~\cite{zhang2011inferring, maathuis2010predicting}. Furthermore, a number of causal structure learning algorithms, for instance, FCI and its variants such as RFCI~\cite{Spirtes2000, colombo2012learning}, and CCD algorithm~\cite{richardson1996discovery}, use PC algorithm as a subroutine.

PC algorithm starts from a complete undirected graph and removes the edges in consecutive levels based on carefully selected conditional independence tests. However, performing these number of tests might take a few days on a single machine in some gene expression data such as DREAM5-Insilico dataset~\cite{marbach2012wisdom}. Furthermore, the order of performing conditional independence tests may affect the final result. 
Parallel implementations of PC algorithm on multi-core CPUs have been proposed in~\cite{madsen2015parallelisation, madsen2017parallel}. 
In~\cite{colombo2014order}, Colombo and Maathuis proposed a variant of PC algorithm called PC-stable which is order-independent and produces less error compared with the original PC algorithm. The key property of PC-stable is that removing an edge in a level has no effect on performing conditional independence tests of other edges in that level. This order-independent property makes PC-stable suitable for executing on multi-core machines.  
In~\cite{le2015fast}, Le et al. proposed a parallel implementation of PC-stable algorithm on multi-core CPUs, called Parallel-PC, which reduces the runtime by an order of magnitude. For instance, it takes a couple of hours to process DREAM5-Insilico dataset. 
In case of using GPU hardware, there was an attempt for parallelization of the PC-stable algorithm in~\cite{schmidt2018order}. However, only a small part (only level zero and level one) of the PC-stable algorithm is parallelized in this method, and thus, it cannot be used as a complete solution in many datasets which require more than two levels. In fact, their approach cannot be generalized to level two and beyond. 

In this paper, we propose a GPU-based parallel algorithm, called ``cuPC", for learning causal structures based on PC-stable. 
We assume that there is no missing observations for any variable, and data has multivariate normal distribution.  
In order to execute PC-stable, one needs to perform conditional independence tests to evaluate whether two variables $V_i$ and $V_j$ are independent given another set of variables $S$. 
The proposed algorithm has two variants, called ``cuPC-E" and ``cuPC-S", which employ the following ideas.

\textit{I)} cuPC-E employs two degrees of parallelism at the same time. First is performing tests for multiple edges in parallel and second, is parallelizing the tests which are performed for a given edge. Although abundant parallelism is available, parallelizing all such tests does not yield the highest performance because it incurs different overheads and also results in many unnecessary tests. 
Instead, cuPC-E judiciously strikes a balance between the two degrees of parallelism in order to efficiently utilize the parallel computing capabilities of GPU and avoid launching unnecessary tests at the same time. In addition, cuPC-E employs two configuration parameters which can be adjusted to tune the performance and achieve high speedup in both sparse and dense graphs.

\textit{II)} A conditional set $S$ might be common in tests of many pairs of variables. cuPC-S takes advantage of this property and reuses the results of computations in one of such tests in the others. 
This sharing can be performed in different ways. For instance, sharing all redundant computations in processing the entire graph might first seem more beneficial, but it has non-justifiable overheads. Hence, cuPC-S employs a carefully-designed local sharing strategy in order to avoid different overheads and achieve significant speedup. 

\textit{III)} cuPC-E and cuPC-S parallel algorithms avoid storing the indices of variables in set $S$. Instead, a combination function is employed to compute the indices on-the-fly and also in parallel.  
%
\textit{IV)} The causal structure is represented by an adjacency matrix which is compacted before starting the computations in every level. The compacted format is judiciously selected to assign and execute parallel threads more efficiently, and also, improves cache performance.  
\textit{V)} GPU shared memory is used in order to improve performance. 
\textit{VI)} Where applicable, threads are terminated early in order to avoid performing unnecessary computations. For instance, edge removals are monitored in parallel, and when an edge is removed in another thread or another block, the rest of the tests on that edge are skipped.

Experiments on multiple datasets show the scalability of the proposed parallel algorithms with respect to the number of variables, the number of samples, and different graph densities. 
For instance, in one of the most challenging datasets, cuPC-S can reduce the runtime of PC-stable from more than $11$ hours to about $4$ seconds. 
On average, cuPC-E and cuPC-S achieve about $500$~X and $1300$~X speedup, respectively, compared to serial implementation on CPU.  

The rest of this paper is organized as follows. In \mRefSec{sec:prelim}, we review some preliminaries on causal Bayesian networks and description of PC-stable. In \mRefSec{sec:alg}, we present the two variants of cuPC algorithm, cuPC-E and cuPC-S. Furthermore, we elaborate details of our contributions in \mRefSec{sec:alg2}. We conduct experiments to evaluate the performance and scalability of the proposed solution in \mRefSec{sec:exp} and conclude our results in \mRefSec{sec:conc}.

\section{Preliminaries}
\label{sec:prelim}

\subsection{Bayesian Networks}
\label{sec:prelim:bn}

Consider a set of \textbf{random variables} $\mSetVar =\{V_1, V_2, \dots, V_n\}$. Given $X, Y, Z \subseteq \mSetVar$, a \textbf{conditional independence (CI) assertion} of the form $X\independent Y | Z$ means $X$ and $Y$ are independent given $Z$. 
A \textbf{CI test} of the form $\mCITest(X,Y | Z)$ is a test procedure based on observed data samples from $X$, $Y$ and $Z$ which determines whether the corresponding CI assertion $X\independent Y | Z$ holds or not. Section~\ref{sec:alg2:citest} describes how to perform CI tests from observed-data samples.  

\textbf{Graphical model} $G$ is a graph which encodes a \textbf{joint distribution} $P$ over the random variables in $\mSetVar$. 
The reason behind the development of a graphical model is that the explicit representation of the joint distribution becomes infeasible as the number of variables grows.
Furthermore, under some assumptions on the data generating model, one can interpret causal relations among the variables from these graphs \cite{peters2017elements}.

\textbf{Bayesian Networks (BN)} are a class of graphical models that represent a factorization of $P$ over $\mSetVar$ by a directed acyclic graph (DAG) $G = (\mSetVar , \mathcal{E})$ as 
\begin{equation}
P(V_1, V_2, \dots, V_n) = \prod_{i = 1}^{n} P(V_i|par(V_i)), 
\label{eq:bn}
\end{equation}
where $\mathcal{E}$ is the set of edges, and $par(V_i)$ denotes parents of $V_i$ in $G$. Moreover, the graph $G$ encodes conditional independence between the random variables in $\mSetVar$ by some notion of separation in graphs. 

A \textbf{Causal Bayesian Network (CBN)} is a BN where each directed edge represents a cause-effect relationship from the parent to its child. For the exact definition of CBN, please refer to \cite{pearl2009causality}, Section $1.3$. A CBN satisfies causal Markov condition, i.e., given $par(V_i)$, variable $V_i$ is independent of any variable $V_j$ that there is no directed path from $V_i$ to $V_j$. 
Let $\mCIAsser(P)$ be the set of all CI assertions that holds in $P$. 
Under causal Markov condition and faithful assumptions~\cite{Spirtes2000}, all CI assertions in $\mCIAsser(P)$ are encoded in the true causal graph $G$ \cite{koller2009probabilistic}.

\subsection{CPDAG}

In a directed graph $G$, we say that three variables $V_i,V_k,V_j \in \mSetVar$ form a \textbf{v-structure} at $V_k$ if variables $V_i$ and $V_j$ have an outgoing edge to variable $V_k$ while they are not connected by any edge in $G$. This is denoted by $V_i \rightarrow V_k  \leftarrow V_j$. 
The \textbf{skeleton} of a directed graph $G$ is an undirected graph that contains edges of $G$ without considering their orientations. 

For a given joint distribution $P$, there might be different DAGs that can represent $\mCIAsser(P)$. The set of all such DAGs is called \textbf{Markov equivalence class}~\cite{parviainen2017learning}. 
It can be shown that two DAGs are in the same Markov equivalence class if they have the same \textbf{skeleton} and the same set of \textbf{v-structures} \cite{andersson1997characterization}. 
A Markov equivalence class can be represented uniquely by a mixed graph called \textbf{completed partial DAG (CPDAG)}. In particular, there is a directed edge in CPDAG from $V_i$ to $V_j$ if this edge exists with the same direction in all DAGs in the Markov equivalent class. There is an undirected edge between $V_i$ and $V_j$ in CPDAG if there exist two DAGs in the Markov equivalence class which have an edge between $V_i$ and $V_j$ but with different orientations.

\subsection{Causal Structure Learning}
\label{sec:prelim:csl}

Causal structure learning, our focus in this paper, is the problem of finding a CPDAG which best describes dependency relations in a given data that is sampled from the random variables in $\mSetVar$. In the literature, two main approaches have been proposed for causal structure learning~\cite{bookci}: constraint-based approach and score-based approach. 

In the constraint-based approach, CI tests are utilized to recover the CPDAG. Examples include PC~\cite{Spirtes2000}, Rank PC~\cite{harris2013pc}, PC-stable~\cite{colombo2014order}, IC~\cite{udea1991equivalence}, and FCI~\cite{Spirtes2000}. 
In the score-based approach, a score function indicates how well each DAG explains dependency relations in the data. Then, a CPDAG with the highest score is obtained by searching over Markov equivalence classes. Examples include Chow-Liu \cite{chow1968approximating} and GES  \cite{chickering2002optimal} algorithms. 
There are other methods such as LiNGAM~\cite{shimizu2006linear, hoyer2008estimation}, and BACKSHIFT~\cite{rothenhausler2015backshift} which do not belong to any of the above two categories because their underlying assumptions are more restricted or their settings are different.

Choosing between the types of algorithms depends on the characteristics of the data \cite{heinze2018causal}. For instance, Scutari et al. \cite{scutari2018learns} concluded that constraint-based algorithms are more accurate than score-based algorithms for small sample sizes and that they are as accurate as hybrid algorithms. 
PC algorithm, as one of the main constraint-based algorithms, has become a common tool for causal explorations and is available in different graphical model learning packages~\cite{kalisch2012causal, scutari2009learning, tetrad}. 
In addition, a number of causal structure learning algorithms utilize PC algorithm as a subroutine~\cite{Spirtes2000,  colombo2012learning, richardson1996discovery}.

\subsection{PC-stable Algorithm}
\label{sec:prelim:stablepc}

In the constraint-based approach, a naive solution to check whether there is an edge between two variables $V_i$ and $V_j$ in the CPDAG is to perform all CI tests of the form $\mCITest(V_i,V_j | S)$ where $S \subseteq \mSetVar\backslash \{V_i, V_j\}$. This solution is computationally infeasible for large number of variables due to exponentially growing number of CI tests.

Unlike the naive solution, the PC algorithm is computationally efficient for sparse graphs with up to thousands number of variables and is commonly used in high-dimensional settings~\cite{kalisch2007estimating}. 
Here we describe PC-stable algorithm which is a variation of PC with less estimation errors~\cite{colombo2014order}. 

PC-stable algorithm consists of two main steps: In the first step, the skeleton is determined by performing a number of carefully selected CI tests. In the second step, the set of v-structures are extracted and as many of the undirected edges as possible are oriented by applying a set of rules called Meek rules~\cite{meek1995causal}. 
The second step is fairly fast. The first step is computationally intensive~\cite{madsen2017parallel} and forms our focus in this paper. 
For instance, in ground truth graphs with a bound $\Delta$ on the maximum degree, the time complexity of PC-stable algorithm is in the order of $O(n^{\Delta})$. 
Sections \ref{sec:alg} and \ref{sec:alg2} present our proposed solution for acceleration of this step on GPU. 
Details of the first step are described in the following. 

\begin{algorithm}[tp]
	\begin{algorithmic}[1]
		\Input $\mSetVar$
		\Output $G$, $\mStoredSepSet$
		\State $G=$ fully connected graph
		\State $\mStoredSepSet = \emptyset$ 
		\State $\mL = 0$ 
		\Repeat
		\State Copy $G$ into $G'$
		\For {any edge $(V_i,V_j)$ in $G$}
		\Repeat
		\State Choose a new $S\subseteq adj(V_i,G') \backslash\{V_j\}$ with $|S| = \mL$ 
		\State Perform $\mCITest(V_i,V_j | S)$
		\If {$V_i \independent V_j|S$}
		\State Remove $(V_i,V_j)$ from $G$
		\State Store $S$ in $\mStoredSepSet$ 
		\EndIf
		\Until{$(V_i,V_j)$ is removed or all sets $S$ are considered}
		\EndFor
		\State $\mL = \mL + 1$ 
		\Until{$($ max degree $-1 \geq \mL~)$}
	\end{algorithmic}
	\caption{The first step in PC-stable algorithm.} 
	\label{alg:spc}
\end{algorithm}

\begin{figure}[tp]
	\centering
	\includegraphics[width =0.8\columnwidth]{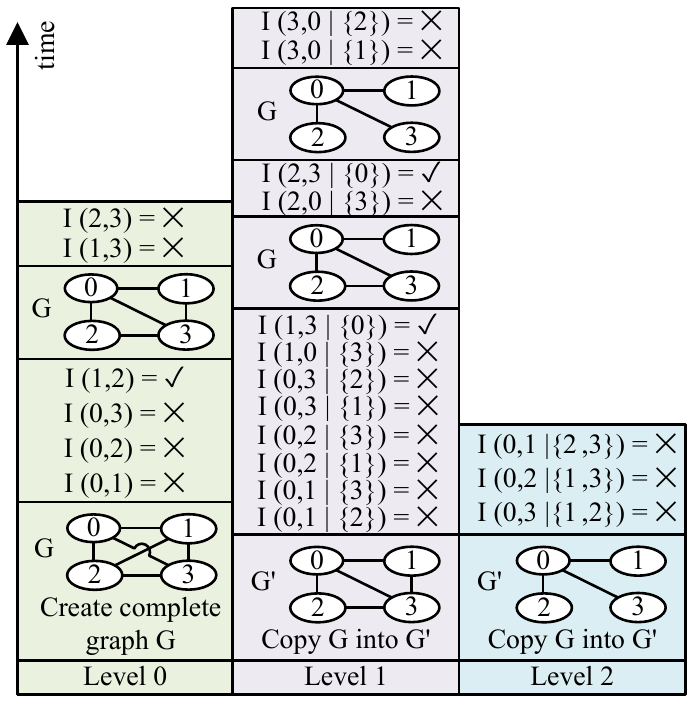}
	\vskip -3mm
	\caption{An example of execution of PC-stable algorithm. 
		For better readability, we use the term $i$ instead of $V_i$. For instance, $\mCITest(0,1)$ actually means $\mCITest(V_0,V_1)$. 
		This is done in \mRefFig{fig:cuPC:E} and \mRefFig{fig:cuPC:S} as well. 
	}
	\label{fig:PC_PCStable_Diagram}	
\end{figure}

See \mRefAlg{alg:spc}. First, $G$ is initiated with a fully connected undirected graph over set $\mSetVar$ (line $1$). 
Next, the extra edges are removed from $G$ by performing a number of CI tests. The tests are performed by levels. In each level $\mL$, first a copy of $G$ is stored in $G'$ (line $5$). Next, for every edge $(V_i,V_j)$ in graph $G$, a CI test $\mCITest(V_i, V_j | S)$ is performed for any $S\subseteq adj(V_i,G') \backslash \{V_j\}$ such that $|S| = \mL$ (lines $6-9$), where $adj(V_i,G')$ denotes the neighbors of $V_i$ in $G'$ (see Section \ref{sec:alg2:citest} for the details of performing a CI test).   
If there exists a set $S$ where $V_i$ is independent of $V_j$ given $S$ (line $10$), edge $(V_i,V_j)$ is removed from $G$, and $S$ is stored in $\mStoredSepSet$ (lines $11-12$). 
Once all the edges are considered, $\mL$ is incremented (line $16$) and the above procedure is repeated. The algorithm continues as long as the maximum degree of the graph is large enough  (line $17$). 
The second step in PC-stable is to use $SepSet$ to find v-structures and orient the edges of graph $G$. 

\mRefFig{fig:PC_PCStable_Diagram} illustrates execution of the first step on a small graph. In level $\mL = 0$, six CI tests are performed, one for every edge in the fully connected graph. Assuming that the result of the fourth CI test is true, we have $V_1 \independent V_2$, and hence, edge $(V_1,V_2)$ is removed. 
In level $\mL = 1$, $12$ CI tests are performed and edges $(V_1,V_3)$ and $(V_2,V_3)$ are removed.

Note that by selecting the conditional sets $S$ from $G'$ but removing edges from $G$, the algorithm finally reaches the same graph regardless of the edge selection order. 
In other words, during the execution of the algorithm in a level, $S$ only depends on $G'$.  
Since performing CI tests in each level is independent of the edge selection order, making an error in one of the CI tests does not have any impact on other CI tests in that level.

\section{cuPC: CUDA-Accelerated PC Algorithm}
\label{sec:alg}

This section presents our proposed solution for acceleration of the computationally-intensive portion of PC-stable (lines $5-15$ in \mRefAlg{alg:spc}) on GPU using CUDA parallel programming API. 
We assume that there is no missing observations for any variable, and data has multivariate normal distribution.  
The overall view of the proposed method is shown in \mRefAlg{alg:gpu}. The main loop on $\mL$ which iterates through the levels still exists in the proposed solution, but the internal computations of every level are accelerated on GPU. 
In specific, since the computations of level zero can be simplified, a separate parallel algorithm is employed for this level (line $7$ in \mRefAlg{alg:gpu}). 
For every level $\mL \geq 1$, first $G$ is copied into $G'$ (line $9$), and then, the required computations are performed (line $10$).  
Note that we work on adjacency matrix of graph $G$ denoted as $A_G$. 
In order to increase the efficiency of the proposed parallel algorithms, $A'_G$ is a compacted version of $A_G$. 
The details are discussed in the following.

A short background on CUDA is presented in \mRefSec{sec:cuda}. Acceleration of level $\mL = 0$ is discussed in \mRefSec{sec:alg:0}.  For levels $\mL \geq 1$, two different parallel algorithms called cuPC-E and cuPC-S are proposed. cuPC-E and the compact procedure are discussed in \mRefSec{sec:alg:v}. cuPC-S is discussed in \mRefSec{sec:alg:s}. Further details on some parts of the proposed solution are discussed later in \mRefSec{sec:alg2}.

\begin{algorithm}[tp]
	\begin{algorithmic}[1]
		\Input $\mSetVar$
		\Output $G$, $\mStoredSepSet$
		\State $G=$ fully connected graph
		\State $\mStoredSepSet = \emptyset$ 
		\State $\mL = 0$ 
		\State $A_G=$ adjacency matrix of graph $G$
		\Repeat
		\If {$(\mL == 0)$} 
		\State GPU: execute level zero
		\Else
		\State GPU: compact $A_G$ into $A'_G$
		\State GPU: execute level $\mL$
		\EndIf	
		\State $\mL = \mL + 1$ 
		\Until{$($ max degree $-1 \geq \mL~)$}
	\end{algorithmic}
	\caption{Overall view of the proposed solution. Lines $7$, $9$, and $10$ are executed in parallel on GPU.} 
	\label{alg:gpu}
\end{algorithm}

\subsection{CUDA}
\label{sec:cuda}

CUDA is a parallel programming API for Nvidia GPUs. GPU is a massively parallel processor with hundreds to thousands of cores. 
CUDA follows a hierarchical programming model. At the top level, computationally intensive functions are specified by the programmer as CUDA \textbf{kernels}. A kernel is specified as a sequential function for a single \textbf{thread}. The kernel is then launched for parallel execution on the GPU by specifying the number of concurrent threads. 

Threads are grouped into \textbf{blocks}. A kernel consists of a number of blocks, and every block consists of a number of threads. 
Every block has access to a small, on-chip and low-latency memory, called \textbf{shared memory}. The shared memory of a block is accessible to all threads within that block, but not to any thread from other blocks\footnote{
	This article is presented based on CUDA programming framework. However, the presented ideas and parallel algorithms can readily be ported to OpenCL programming framework for other GPU vendors as well. In specific, block, thread and shared memory in CUDA programming framework correspond to work-group, work-item, and local memory in OpenCL programming framework.
}.

In order to identify blocks within a kernel, and also, threads within a block, a set of indices are used in the CUDA API, for instance, $blockIdx.y$ and $blockIdx.x$ as the block index in dimension $y$ and dimension $x$ within a 2D kernel, and $threadIdx.y$ and $threadIdx.x$ as the thread index in dimensions $y$ and $x$ within a 2D block. For brevity, we denote these four indices as $by$, $bx$, $ty$ and $tx$, respectively.

\subsection{Level $\mL = 0$}
\label{sec:alg:0}

Size of conditional sets $S$ is equal to $\mL$ (\mRefAlg{alg:spc}, line $8$). As a result, in level zero, $S = \emptyset$, and therefore, the required computations can be simplified. 
In specific, for every edge $(V_i,V_j)$ in $G$, only one CI test is required, which is $\mCITest(V_i,V_j|\emptyset)$ or simply $\mCITest(V_i,V_j)$. In addition, copying $G$ into $G'$ is not required.

All the required CI tests $\mCITest(V_i,V_j)$ are performed in parallel as shown in \mRefAlg{alg:L0}. Since the input graph in level zero is a fully connected undirected graph,  a total of $n(n-1)/2$ tests are required, i.e., one for every edge. Every test $\mCITest(V_i,V_j)$ is assigned to a separate thread and $n^2$ threads are launched. Threads are grouped in a 2D kernel of $\nicefrac{n}{32} \times \nicefrac{n}{32}$ blocks. Every block has $32 \times 32$ threads. Indices $i$ and $j$ are calculated in lines $1-2$. Here, $0 \leq by, bx < \nicefrac{n}{32}$ and $0 \leq ty, tx < 32$. 

Lines $4-7$ are executed in only $n(n-1)/2$ threads. In line $4$, the CI test $\mCITest(V_i,V_j)$ is performed. In lines $5-7$, the edge $(V_i,V_j)$ is removed from graph $G$ if $V_i \independent V_j$. 
The term $A_G$ denotes the adjacency matrix of graph $G$. Edge $(V_i,V_j)$ is removed from $G$ by setting $A_G[i,j]=A_G[j,i]=0$.

\begin{algorithm}[tp] 
	\begin{algorithmic}[1]
		\Input $A_G$ 
		\Output $A_G$
		\Block $\nicefrac{n}{32} \times \nicefrac{n}{32}$
		\Thread $32 \times 32$ 
		\State $i = by \times 32 + ty$ \label{alg:L0:ij_1}
		\State $j = bx \times 32 + tx$ \label{alg:L0:ij_2}
		\If {$(i<j)$} 
		\State Perform $I(V_i,V_j)$ \label{alg:L0:CI}
		\If {$(V_i \independent V_j)$} \label{alg:L0:if}
		\State $A_G[i,j] = A_G[j,i] = 0$ \label{alg:L0:remove}
		\EndIf
		\EndIf
	\end{algorithmic}
	\caption{Acceleration of level $\mL=0$. See Section \ref{sec:alg:0}.} 
	\label{alg:L0}
\end{algorithm}


\begin{figure}[tp]
	\centering
	\includegraphics[width = \columnwidth]{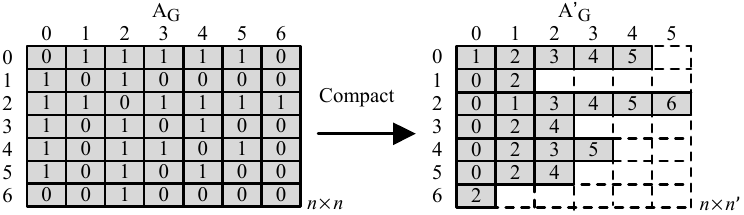}	
	\vskip -3mm
	\caption{$A'_G$ is formed by compacting $A_G$.} 
	\label{fig:cuPC:Compact} 
\end{figure}

\subsection{Level $\mL \geq 1$: Parallel Algorithm cuPC-E}
\label{sec:alg:v}

Two different parallel algorithms (kernels) are proposed for acceleration of every level $\mL \geq 1$. This section describes the first algorithm, called cuPC-E. See \mRefAlg{alg:cuPC:E}.

\textit{Compact:} cuPC-E takes $\mL$, $A_G$ and $A'_G$ as input. As shown in line $9$ in \mRefAlg{alg:gpu}, $A'_G$ is formed by compacting adjacency matrix $A_G$ into a sparse representation. 
\mRefFig{fig:cuPC:Compact} illustrates a small example. An element with value $j$ in $i$-th row in $A'_G$ denotes existence of edge $(V_i,V_j)$ in $A_G$. $A'_G$ has $n$ rows. Row $i$ has $n'_i$ elements, i.e., edges. Let $\displaystyle n' = \max_{0 \leq i < n} n'_i$.  
Note that $A'_G$ can be implemented in different formats such as linked lists as in adjacency list representations \cite{saad1994sparskit}. However, since linked lists are not efficient for parallel execution, $A'_G$ is implemented as a matrix with $n$ rows and $n'+1$ columns. The element at the last column of each row $i$ stores $n'_i$.  
The $Compact$ procedure is executed in parallel by employing another parallel algorithm called $scan$ \cite{hillis1986data, billeter2009efficient}. Details are removed for brevity.

\begin{figure}[tp]
	\centering
	\includegraphics[width = \columnwidth]{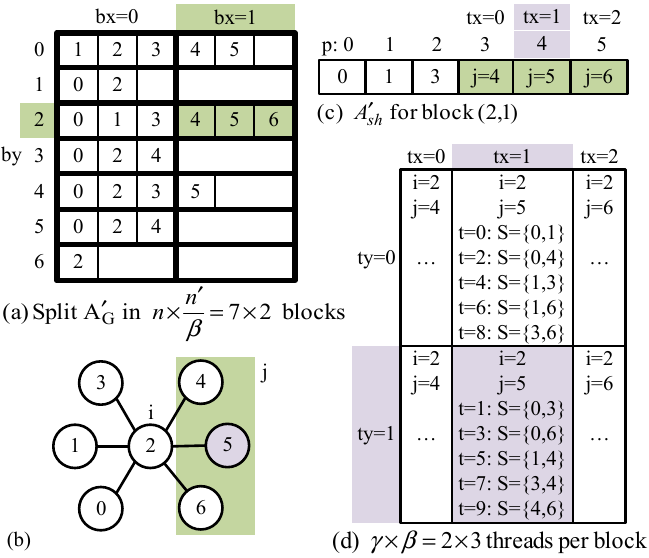}	
	\vskip -3mm
	\caption{Blocks and threads in cuPC-E parallel algorithm. In this example, $n=7$, $n'=6$, $\mEB=3$, $\mET=2$ and $\mL=2$. Block $(2,1)$ is marked with green color, and thread $(1,1)$ in this block is marked with purple color. For threads $(0,1)$ and $(1,1)$ in block $(2,1)$, $S$ is selected from set $\{0,1,3,4,6\}$.}
	\label{fig:cuPC:E}
	\vskip -3mm
\end{figure}

\textit{Blocks and Threads:} cuPC-E kernel consists of $n \times \nicefrac{n'}{\mEB}$ blocks. See \mRefFig{fig:cuPC:E}(a). Every block performs the required CI tests for $\mEB$ edges, i.e., $\mEB$ consecutive elements from one row in $A'_G$. In \mRefFig{fig:cuPC:E}(a), there are $7 \times 2$ blocks. Block $(2,1)$, which is marked with green color, works on $\mEB=3$ edges, namely, $(V_2,V_4)$, $(V_2,V_5)$, and $(V_2,V_6)$. See \mRefFig{fig:cuPC:E}(b). 
The CI tests for each one of the $\mEB$ edges are split among $\mET$ threads. Hence, every block consists of $\mET \times \mEB$ threads. In \mRefFig{fig:cuPC:E}(d), there are $2 \times 3$ threads in block $(2,1)$. Thread $(1,1)$ in this block, which is marked with purple color, works on half of the CI tests for edge $(V_2,V_5)$. Thread $(0,1)$ works on the other half.

\begin{algorithm}[tp]
	\begin{algorithmic}[1]
		\Input $A_G$, $A'_G$, $\mL$ 
		\Output $A_G$, $\mStoredSepSet$
		\Block $n \times \nicefrac{n'}{\mEB}$
		\Thread $\mET \times \mEB$ 
		\State $i = by$ 
		\State $n'_i =$ size of row $i$ in $A'_G$ 
		\State Copy the entire row $i$ from matrix $A'_G$ into vector $\mSh$ in shared memory 
		\State $p = bx \times \mEB + tx$ 
		\State $j = \mSh[p]$ 
		\For{$~~(t=ty;~~ t < \binom{n'_i-1}{\mL};~~ t=t+\mET)$} 
		\If {$(A_G[i,j] == 1)$} \label{alg:cuPC:E:if}
		\State $P_{1 \times \mL} = Comb(n'_i-1, \mL, t, p )$ 
		\State $S_{1 \times \mL} = \mSh[ P ]$ 
		\State Perform $\mCITest(V_i,V_j | S)$ 
		\If {$(V_i \independent V_j | S)$}
		\State $A_G[i,j] = A_G[j,i] = 0$
		\State Store $S$ in $\mStoredSepSet$ 
		\EndIf
		\EndIf
		\EndFor
	\end{algorithmic}
	\caption{Acceleration of level $\mL \geq 1$ with parallel algorithm cuPC-E. See Section \ref{sec:alg:v} and Fig.~\ref{fig:cuPC:E}.} 
	\label{alg:cuPC:E}
\end{algorithm}

\textit{Shared Memory:} The threads in block $(by,bx)$ frequently access different elements in row $by$ in $A'_G$. Therefore, in order to speedup the memory accesses, the entire row is copied into the block's shared memory, i.e., into vector $\mSh$. See Fig.~\ref{fig:cuPC:E}(c). 

\textit{Index Calculations:} Let $(V_i,V_j)$ denote the target edges in block $(by,bx)$. For all threads within block $(by,bx)$, $i$ is equal to $by$. See line $1$ in \mRefAlg{alg:cuPC:E}. For thread $(ty,tx)$ in this block, $j$ is equal to the $tx$-th element in the green portion of the corresponding row, i.e., element $bx \times \mEB + tx$ in $\mSh$. See lines $4-5$ in \mRefAlg{alg:cuPC:E}, and also, Fig.~\ref{fig:cuPC:E}(c). 

\textit{Combinations:} Consider all CI tests $\mCITest(V_i,V_j | S)$ for edge $(V_i,V_j)$. Set $S$ is formed by selecting $\mL$ elements from row $i$ in $A'_G$ or equivalently from $\mSh$. Since element $j$ in this row should not be selected, there will remain $n'_i-1$ elements to choose from. Therefore, there are a total of $\binom{n'_i-1}{\mL}$ possible combinations for set $S$. CI tests of an edge $(V_i,V_j)$ are split among $\mET$ threads. 
In the example of Fig.~\ref{fig:cuPC:E}(d), $i=2$ and $j=5$. Hence, $S$ is selected from $\{0,1,3,4,6\}$. There are $\binom{5}{2}=10$ possible combinations for $S$. Each of the $\mET=2$ threads sequentially perform $\nicefrac{10}{2}=5$ of these tests. 
See lines $6-10$ in \mRefAlg{alg:cuPC:E}, and also, Fig.~\ref{fig:cuPC:E}(d). $P$ is an array of pointers that point to the selected elements. For instance, when $t=9$ (the last combination), we have $P=\{3,5\}$ and $S=\{V_4,V_6\}$.  
The $\mCombFunc$ function returns $t$-th combination in parallel, while skipping the unwanted combinations that include $p$, i.e., the pointer to $j$. Different parallel threads call this function with different values of $t$. The internal details of the $\mCombFunc$ function are discussed later in \mRefSec{sec:alg2:comb}.  

\textit{Edge Removal:} In lines $10-14$, one CI test $\mCITest(V_i,V_j | S)$ is performed, and if $V_i \independent V_j | S$, the edge $(V_i,V_j)$ is removed from $A_G$. As mentioned before in \mRefAlg{alg:spc}, the conditional sets $S$ are selected from $G'$ but the edges are removed from $G$.

\textit{Key Features:} Important features of cuPC-E parallel algorithm are discussed in the following.  
\textit{I)} cuPC-E offers two degrees of parallelism, in specific, processing all the edges in parallel, and for every edge, performing the CI tests in parallel. 
Although abundant parallelism is available, parallelizing all such tests does not yield the highest performance. cuPC-E does not fully parallelize all CI tests for an edge. The number of CI tests for edge $(V_i,V_j)$ is equal to $\binom{n'_i-1}{\mL}$, while these CI tests are performed by only $\mET$ parallel threads. In the example of \mRefFig{fig:cuPC:E}(d), for edge $(V_2,V_5)$, $10$ tests are performed by $2$ parallel threads. 
When one of these $\mET$ threads removes the target edge, we no longer need to perform the rest of the CI tests for that edge. The $if$ statement in line $7$ in \mRefAlg{alg:cuPC:E} blocks these unnecessary tests. 
$\mET=1$ avoids all the unnecessary tests but is sequential, and $\mET=\binom{n'_i-1}{\mL}$ is fully parallel but does not avoid any of the unnecessary tests. Parallel algorithm cuPC-E, therefore, strikes a balance by judiciously employing partial parallelism of the CI tests.

\textit{II)} Edge removals are monitored in parallel in order to avoid unnecessary tests. In specific, when edge $(V_i,V_j)$ is removed by another block, i.e., by a block with $by=j$, the same $if$ statement in line $7$ in \mRefAlg{alg:cuPC:E} blocks the unnecessary tests.

\textit{III)} All indices required for fetching sets $S$ are calculated on-the-fly and also in parallel based on a combination function (\mRefSec{sec:alg2:comb}), and hence, cuPC-E does not use extra memory for storing the indices. 

\textit{IV)} Processing the compacted version of the adjacency matrix removes unnecessary checks for zero elements of $A_G$, reduces total number of combinations for set $S$, and also leads to better cache performance.  The compacted format is judiciously selected to match the proposed method.

\textit{V)} Use of shared memory for the rows of $A'_G$ increases the performance. Note that every block has only one copy of its corresponding row but processes $\mEB$ edges. 
Storing the correlation matrix $C$ or the set of combinations in shared memory is not beneficial.

\subsection{Level $\mL \geq 1$: Parallel Algorithm cuPC-S}
\label{sec:alg:s}

Every CI test $\mCITest(V_i,V_j | S)$ includes computing pseudo-inverse of a matrix $M_2$. See Sections \ref{sec:alg2:citest} and \ref{sec:alg2:inverse} for the details. Pseudo-inverse computations are time consuming. 
cuPC-S employs the following idea in order to accelerate the process. The matrix $M_2$, which requires inversion, depends only on set $S$, and not $V_i$ or $V_j$. See \mRefEq{eq:PartialCorMat}. 
Therefore, by assigning the CI tests that depend on the same set $S$ to a single thread, it is possible to avoid multiple calculations of the same pseudo-inverse by sharing it among the CI tests. 
See \mRefAlg{alg:cuPC:S}. 

\vskip 1mm
\textit{Blocks and Threads:} cuPC-S kernel consists of $n\times\mSB$ blocks. 
For a given row $i$ in $A'_G$, there exist $\binom{n'_i}{\mL}$ possible sets $S$ of size $\mL$. Processing of these sets are split among $\mSB$ blocks, each containing $\mST$ threads. Each one of these $\mSB \times \mST$ threads, therefore, is responsible for processing $\binom{n'_i}{\mL} / (\mSB \times \mST)$ sets $S$. 

\begin{figure}[tp]
	\centering
	\includegraphics[width = 0.95\columnwidth]{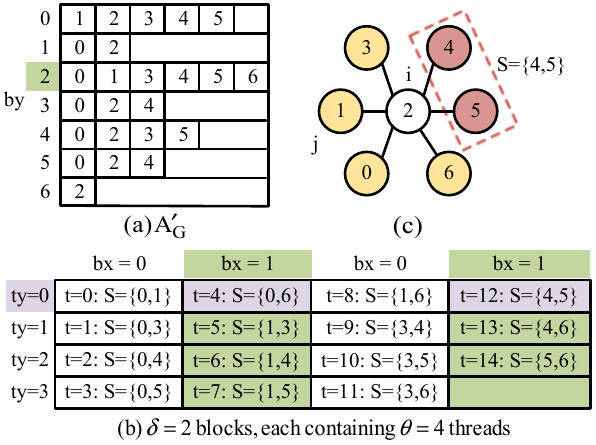}	
	\vskip -3mm
	\caption{
		Blocks and threads in cuPC-S parallel algorithm. In this example, $n=7$, $n'=6$, $\mSB=2$, $\mST=4$ and $\mL=2$. Block $(2,1)$ is marked with green color, and thread $0$ in this block is marked with purple color. 
		In the second loop iteration in this thread, we have $t=12$, and hence, $S=\{4,5\}$ (red color). Therefore, $j$ is equal to $0$, $1$, $3$, and finally $6$ (orange color). 
	}
	\label{fig:cuPC:S}
\end{figure}

\mRefFig{fig:cuPC:S} illustrates a small example. Row $2$ contains $n'_2=6$ elements. Therefore, there are $\binom{6}{2}=15$ possible sets $S$ for this row, which are split among $\mSB=2$ blocks, each containing $\mST=4$ threads. See \mRefFig{fig:cuPC:S}(b). 
Block $(2,1)$ is marked with green color, and thread $0$ within this block is marked with purple color. 
This thread works on two sets $S$, in specific, $S=\{V_0,V_6\}$ and $S=\{V_4,V_5\}$.

\begin{algorithm}[tp]
	\begin{algorithmic}[1]
		\Input $A_G$, $A'_G$, $\mL$ 
		\Output $A_G$, $\mStoredSepSet$
		\Block $ n \times \mSB$ 
		\Thread $ \mST \times 1$  
		\State $i = by$ 
		\State $n'_i =$ size of row $i$ in $A'_G$ 
		\State Copy the entire row $i$ from matrix $A'_G$ into vector $\mSh$ in shared memory  
		\For{$~~(t = bx \times \mST + ty;~~t<\binom{n'_i}{\mL};~~t=t+ \mST \times \mSB)$}  
		\State $P_{1 \times \mL} = Comb(n', \mL, t)$ 
		\State $S_{1 \times \mL} = A'_{sh} [ P ]$ 
		\State Form matrix $M_2$ based on set $S$ (Section~\ref{sec:alg2:citest})
		\State $M^{-1}_2 = $ Pseudo-inverse of $M_2$ (Section~\ref{sec:alg2:inverse})
		\For{$~~p = 0~~$ to $~~n'_i~~$} 
		\State $j = \mSh[p]$ 
		\If {$(j \notin S)$} 
		\If {$(A_G[i,j] == 1)$} \label{alg:cuPC:S:if}
		\State Perform $\mCITest(V_i,V_j | S)$ 
		\If {$(V_i \independent V_j | S)$}
		\State $A_G[i,j] = A_G[j,i] = 0$
		\State Store $S$ in $\mStoredSepSet$
		\EndIf
		\EndIf
		\EndIf
		\EndFor
		\EndFor
	\end{algorithmic}
	\caption{Acceleration of level $\mL \geq 1$ with parallel algorithm cuPC-S. See Section \ref{sec:alg:s} and Fig.~\ref{fig:cuPC:S}.} 
	\label{alg:cuPC:S}
\end{algorithm}

\vskip 1mm
\textit{Index Calculations:} Lines $1-3$ in \mRefAlg{alg:cuPC:S} are similar to cuPC-E. 
Since every thread that is assigned to row $i=by$ in cuPC-S is responsible for processing $\binom{n'_i}{\mL} / (\mSB \times \mST)$ sets $S$, the $for$ loop in line $4$ iterates $\binom{n'_i}{\mL} / (\mSB \times \mST)$ times. 
In \mRefFig{fig:cuPC:S}(b), it iterates twice, for instance, we have $t=1\times4+0=4$ and $t=4+8=12$ in thread $0$ in block $(2,1)$.

In every iteration, one set $S$ is selected based on the value of $t$. This is done using the $Comb$ function. See lines $5-6$ in \mRefAlg{alg:cuPC:S}. 
The selected set $S$ is used to perform a number of CI tests $\mCITest(V_i,V_j | S)$. Since matrix $M_2$ depends only on $S$, and not $V_i$ or $V_j$, we compute this matrix and its pseudo-inverse once and use the results in all these CI tests. See lines $7-8$.

In the target CI tests $\mCITest(V_i,V_j | S)$, $i=by$ and different values of $j$ are determined in lines $9-11$ by iterating through all adjacent nodes of $V_i$ and selecting the ones which are not in $S$. 
As an example, consider thread $0$ in block $(2,1)$ in \mRefFig{fig:cuPC:S}. This thread has two loop iterations: $t=4$ and $t=12$. In the second iteration ($t=12$), we have $S=\{V_4,V_5\}$ which is marked with red color in the figure. As a result, $V_j$'s are the other adjacent nodes of $V_i$, namely, $V_0$, $V_1$, $V_3$ and finally $V_6$. They are marked with orange color. See \mRefFig{fig:cuPC:S}(c). 
Hence, in the second iteration ($t=12$) in thread $0$ in block $(2,1)$, the following CI tests are performed: 
$\mCITest(V_2,V_0 | \{V_4,V_5\})$, 
$\mCITest(V_2,V_1 | \{V_4,V_5\})$, 
$\mCITest(V_2,V_3 | \{V_4,V_5\})$, and 
$\mCITest(V_2,V_6 | \{V_4,V_5\})$.

\vskip 2mm
\textit{Edge Removal:} Lines $12-18$ in \mRefAlg{alg:cuPC:S} are similar to cuPC-E, except that line $13$ executes faster because part of performing a CI test is to compute pseudo-inverse $M^{-1}_2$ which is already computed in line $8$.

\vskip 2mm
\textit{Key Features:} Similar to cuPC-E parallel algorithm, cuPC-S \textit{I)} works on $A'_G$ which is the compacted version of the adjacency matrix, \textit{II)} employs shared memory, \textit{III)} skips unnecessary CI tests via the $if$ statement in line $12$ in \mRefAlg{alg:cuPC:S}, and \textit{IV)} employs a parallel combination function to compute the indices of sets $S$. 
\textit{V)} More importantly, sharing one pseudo-inverse among multiple CI tests brings a large saving.  

\textit{VI)} In the CUDA framework, every $32$ threads within a block form a warp. Therefore, in order to maximize GPU utilization, the number of threads within a block, i.e., $\mST$, should be a multiple of $32$. However, $\binom{n'_i}{\mL}$ might not be divisible by $\mSB \times \mST$. 
cuPC-S employs the following idea in order to resolve this issue.  
Blocks do not process all their assigned sets $S$ in parallel. Instead, they iterate multiple times and in every iteration, process $\mST$ sets $S$, where $\mST$ is a multiple of $32$. 
As a result, only the last iteration may not contain a multiple of $32$ active threads.

\textit{VII)} There are many CI tests $\mCITest(V_i,V_j | S)$ that share the same set $S$. For instance, in \mRefFig{fig:cuPC:S}, $S=\{V_4,V_5\}$ can be shared among CI tests in not only row $2$ but also row $0$ because both of these rows have elements $4$ and $5$, i.e., because both $V_0$ and $V_2$ are connected to $V_4$ and $V_5$. See \mRefFig{fig:cuPC:S}(a). 
cuPC-S only shares a set $S$ and its corresponding pseudo-inverse $M^{-1}_2$ locally. In other words, a set $S$ is shared only among the CI tests $\mCITest(V_i,V_j | S)$ with the same value $i$, i.e., among the CI tests of edges which are connected to the same $V_i$. 
This is in contrast to sharing a set $S$ globally, i.e., among all CI tests from the entire graph. 
While global sharing may yield more savings, it requires searching the entire graph. The amount of extra saving is not large enough to justify the additional cost of global search. 
\mRefSec{sec:exp:GlobalVsSemi} demonstrates this point through an experiment.

\section{Further Details of cuPC}
\label{sec:alg2}

\subsection{Early Termination}
\label{sec:alg2:early}

So far we have discussed only one of the early termination strategies employed in cuPC, in specific, the $if$ statements in line $7$ in \mRefAlg{alg:cuPC:E}, and line $12$ in \mRefAlg{alg:cuPC:S}. 
There are other cases where further processing is no longer required, and threads may terminate early in order to save time. 
Such cases are listed in the following. For brevity, their corresponding $if$ statements are not shown in \mRefAlg{alg:cuPC:E} and \mRefAlg{alg:cuPC:S}. 
\textit{I)} If the number of adjacent nodes of $V_i$ is less than $\mL+1$, i.e., $n'_i < \mL+1$, then all threads in the corresponding blocks are terminated because we need at least one adjacent node $V_j$ plus $\mL$ other adjacent nodes for set $S$. 
\textit{II)} In block $(by,bx)$ in cuPC-E, if $bx \times \mEB \geq n'_i$, all threads terminate. This is because $n'_i$, i.e., the number of edges to be processed in row $i=by$, is too small to require the processing power of this block. 
\textit{III)} Similarly, in block $(by,bx)$ in cuPC-S, if $bx \times \mST \geq \binom{n'_i}{\mL}$, all threads terminate. This is because $\binom{n'_i}{\mL}$, i.e., the number of sets $S$ in row $i=by$, is too small.

\subsection{Computing Sets of Combination in Parallel}
\label{sec:alg2:comb}

The $\mCombFunc$ function employed in \mRefAlg{alg:cuPC:E} and \mRefAlg{alg:cuPC:S} is discussed in this section. 
Let $O=\{O_0,O_1, O_2, \cdots, O_{{n \choose \mL} -1}\}$ be the set of all possible combinations of choosing $\mL$ elements from set $\{1, 2, 3, \cdots, n\}$ in lexicographical order. For instance, when $n = 3$ and $\mL = 2$, we have $O_0=[1,2]$, $O_1=[1,3]$, and $O_2=[2,3]$. 
Given $n$, $\mL$ and $t$, the algorithm in~\cite{buckles1977algorithm} directly computes vector $O_t$ without requiring to compute the entire set $O$. Thus, by utilizing this algorithm in every thread, every $O_t$ is derived separately.

There are $\mL$ elements in $O_t$. Let $O_t=[O_t[0], \dots, O_t[\mL-1]]$ and $O_t[-1] = 0$. According to \cite{buckles1977algorithm}, the following statement holds true: 
\begin{equation} 
t  = \sum_{c = 0}^{\mL-1} 
\sum_{k = O_t[c - 1] + 1}^{O_t[c] - 1} {n-k \choose \mL - (c+1)} 
\label{eq:comb}
\end{equation}

Based on the above equation, \mRefAlg{alg:ParComb} iteratively computes $O_t$. The algorithm has $\mL$ iterations. In iteration $c$, $O_t[c]$ is computed. The value of $Sum$ must be less than or equal to, and also, as close as possible to the value of $t$. 

Once all the $\mL$ elements in $O_t$ are computed in \mRefAlg{alg:ParComb}, the following minor modifications are performed in order to use the results in cuPC-E and cuPC-S parallel algorithms. 
In cuPC-S, since all indices start from zero (and not one), all elements in $O_t$, i.e., the output of \mRefAlg{alg:ParComb}, are decremented by $1$. 
In cuPC-E, in addition to the above modification, we also need to skip all the combinations which include $p$, i.e., the index of $j$. Hence, we set the input of the $Comb$ function to $n'_i-1$ instead of $n'_i$, and also, increment all the values which are larger than or equal to $p$ by $1$.

\begin{algorithm}[tp]
	\begin{algorithmic}[1]
		\Input $n$, $\mL$, $t$, $p$
		\Output $O_t$ 
		\State $Sum = 0$
		\State $O_t[-1] = 0$
		\For {$c = 0$ \textbf{to} $\mL-1$}
		\State	$O_t[c] = O_t[c-1]$ 
		\While{$Sum \leq t$}
		\State $O_t[c] = O_t[c] + 1$
		\State $Sum = Sum + {{n - O_t[c]} \choose {\mL - (c + 1)}}$
		\EndWhile \label{alg:ParComb_1}
		\State $Sum = Sum - {{n - O_t[c]} \choose {\mL - (c + 1)}}$ \label{alg:ParComb_2}	
		\EndFor
	\end{algorithmic}
	\caption{Combination function.}
	\label{alg:ParComb}
\end{algorithm}

\subsection{CI Tests}
\label{sec:alg2:citest}

In practice, the CI tests need to be performed based on data samples observed from the random variables. 
In particular, for multivariate normal distribution, CI test $\mCITest(V_i,V_j | S)$ can be performed based on partial correlations. 
Let $\rho(V_i,V_j|S)$ be the partial correlation between $V_i$ and $V_j$ given $S$. Then, we have $V_i \independent V_j | S$ if and only if $\rho(V_i,V_j|S)$ is zero. 
The exact procedure is described below: 
Let $C_{n \times n}$ be the correlation matrix among the $n$ random variables in the set $\mSetVar$, and $C[V_i,V_j]$ be $(i,j)$-th entry in matrix $C$. We define a $1 \times \mL$ vector $C(V_i,S)$ as 
\begin{equation}
C(V_i,S) := \Big[ C[V_i,S[1]], C[V_i,S[2]],\dots  C[V_i,S[\mL]] \Big]_{1 \times \mL},
\end{equation}
where $S[k]$ is the $k$-th element in the set $S$. 
In order to compute $\rho(V_i,V_j|S)$, we first extract $M_0, M_1$, and $M_2$ matrices from the correlation matrix  as the following:

\begin{align}
\nonumber M_0 &= \begin{bmatrix}
C[V_i,V_i] & C[V_i,V_j] \\ 
C[V_j,V_i] & C[V_j,V_j]
\end{bmatrix}_{2 \times 2},\quad
M_1 = \begin{bmatrix}
C(V_i,S)\\ 
C(V_j,S)
\end{bmatrix}_{2 \times \mL},\\ 
M_2 &= \begin{bmatrix}
C(S[1],S) \\ 
C(S[2],S) \\ 
\vdots\\
C(S[\mL],S)
\end{bmatrix}_{\mL \times \mL}.
\label{eq:PartialCorMat}
\end{align}

Next, we obtain matrix $H = M_0 - M_1 \times M_2^{-1} \times M_1^T$. Note that $M_2$ might be ill-conditioned, and hence, $M_2^{-1}$ needs to be computed using a pseudo-inverse algorithm (\mRefSec{sec:alg2:inverse}). 
Once $H$ which is a $2 \times 2$ matrix is computed, an estimation of $\rho(V_i,V_j|S)$ is computed as the following:
\begin{equation}
\hat{\rho}(V_i,V_j|S) = \dfrac{H[1,2]}{ \sqrt{H[1,1] \times H[2,2]} }.
\label{eq:RhoCal}
\end{equation}

In order to test whether the value of $\hat{\rho}(V_i,V_j|S)$ implies  $V_i \independent V_j | S$, we compute Fisher's z-transform \cite{kalisch2007estimating} as 
\begin{equation}
Z(\hat{\rho}(V_i,V_j|S)) = \left| \dfrac{1}{2} \times \ln \left(\dfrac{1 + \hat{\rho}(V_i,V_j|S)}{1 - \hat{\rho}(V_i,V_j|S)}\right)\right|, 
\label{eq:FisherZTransform}
\end{equation}
and compare it with the following threshold:
\begin{equation}
\mTh = \dfrac{\Phi^{-1}(1-\dfrac{\alpha}{2})}{\sqrt{m - |S| - 3}},	
\label{eq:threshold}
\end{equation}
where $m$, $\alpha$ and $\Phi$ are the size of data samples for every random variable, the significance level for testing partial correlations, and CDF of standard normal distribution, respectively. 
If $Z(\hat{\rho}(V_i,V_j|S)) \leq \mTh$, we imply that $V_i \independent V_j | S$. 
Note that in level zero, the above procedure is reduced to comparing $Z(C[V_i,V_j])$ with the threshold $\mTh$. 

We can conclude that a CI test $\mCITest(V_i,V_j | S)$ can be performed based on observational data, in specific, based on the threshold $\mTh$ and the correlation matrix $C_{n \times n}$ among the $n$ random variables. 

\subsection{Pseudo-Inverse}
\label{sec:alg2:inverse}

As mentioned above, a pseudo-inverse algorithm is needed in order to compute $M_2^{-1}$. We employ Moore-Penrose~\cite{PseudoInverse} method as shown in \mRefAlg{alg:pseudoInverse}.  The pseudo-inverse is computed based on two matrices $L$ and $R$. Matrix $L$ is computed as the full rank Cholesky factorization of matrix $M_2^T\times M_2$. Matrix $R$ is computed as the inverse (the usual inverse) of $L^T \times L$.

\begin{algorithm}[tp]
	\caption{Pseudo-inverse method.}
	\label{alg:pseudoInverse}
	\begin{algorithmic}[1]
		\Input $M_2$
		\Output $M_2^{-1}$
		\State $L =~$ Cholesky Factorization $~(M_2^T\times M_2)$
		\State $R = ( L^T \times L)^{-1}$
		\State $M_2^{-1} = L \times R \times R \times L^T \times M_2^T$
	\end{algorithmic}
\end{algorithm}

\section{Experimental Evaluation}
\label{sec:exp}

\subsection{Source Code}
\label{sec:exp:src}

cuPC is implemented in the \texttt{C} language in the CUDA framework.  
Our parallel implementation is wrapped in a function in the \texttt{R} language with the exact same interface as the original PC-stable function in pcalg \cite{kalisch2012causal}. Thus, cuPC is consistent with standard casual learning \texttt{R} packages and can be easily integrated in pcalg. 
The source code of cuPC is available online \cite{sourcecode}.

\begin{table*}[bp]
	\centering
	\renewcommand\thetable{2} 
	\caption{
		Comparing serial, multicore, and GPU implementations. 
		The first five rows show the runtime values, which are denoted as T1 to T5. 
		The last three rows show speedup ratios, which are calculated as T1/T2, T3/T4 and T3/T5. 
		The last column compares the geometric mean of speedup ratios.
	}
	\vskip -3mm
	\includegraphics[width = 1.0\textwidth]{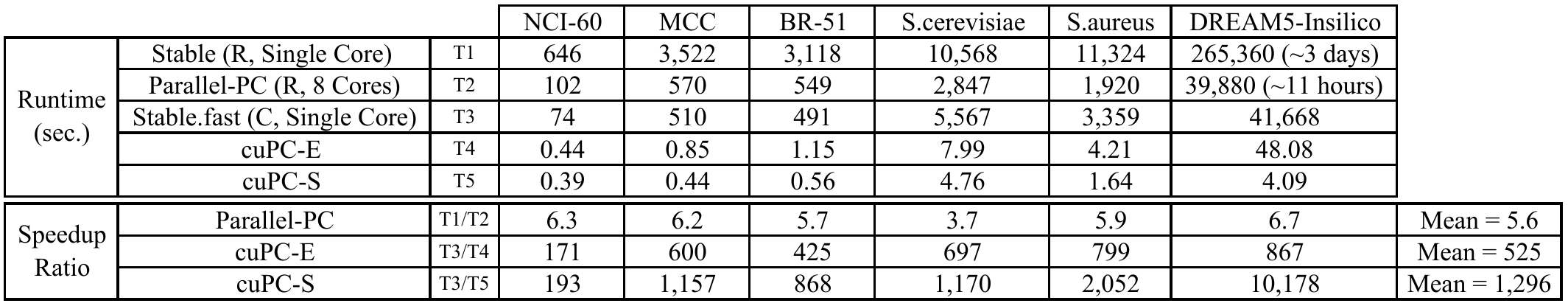}
	\label{tab:overall}
\end{table*}

\subsection{Experiment Setup}
\label{sec:exp:setting}

We experimentally evaluate cuPC along with the following related previous works. 
Two different serial implementations of PC-stable \cite{colombo2014order} algorithm are available as part of the pcalg \cite{pcalg} package. The original one (called "Stable" in pcalg) is implemented in \texttt{R} language, and the recent one (called "Stable.fast") is in \texttt{C} language. 
A multi-threaded method, called "Parallel-PC" \cite{le2015fast}, is implemented in \texttt{R} language and is available here \cite{ParallelPC}. 
In addition, Stable.fast (i.e., the \texttt{C} implementation in pcalg) supports multi-threaded execution mode as well.

We employ a machine with an Intel Xeon CPU with $8$ cores running at $2.5$ GHz. 
Serial methods (Stable and Stable.fast) are executed on a single core, and multi-threaded methods (Parallel-PC and Stable.fast) are executed on all the $8$ cores. 
The CUDA kernels in cuPC are executed on Nvidia GTX $1080$ GPU which is hosted on the same machine, and the other procedures in cuPC are executed sequentially on a single core. 
We employ Ubuntu OS $16.04$, gcc version $5.4$, and CUDA version $9.2$. 

Six gene expression datasets are employed as our benchmarks \cite{le2014mirna, maathuis2010predicting, marbach2012wisdom}. These are the same benchmarks used in \cite{le2015fast}. Table~\ref{tab:datasetSpec} shows the number of random variables and the number of samples in every dataset. 

The accuracy of the proposed method is exactly the same as the one of PC-stable which was evaluated extensively in \cite{colombo2014order} in terms of True Discovery Rate (TDR) and Structural Hamming Distance (SHD). 
This is because cuPC is GPU-accelerated implementation of the same PC-stable algorithm.

\begin{table}[H]
	\centering
	\renewcommand\thetable{1} 
	\caption{Benchmark datasets.}
	\vskip -3mm
	\label{tab:datasetSpec}
	\begin{tabular}{|c|c|c|}
		\hline
		Dataset 		& \# of variables ($n$)	& \# of samples ($m$)	\\
		\hline
		NCI-60 			& 1190 							& 47			\\
		\hline
		MCC 			& 1380 							& 88			\\
		\hline
		BR-51 			& 1592 							& 50			\\
		\hline
		S.cerevisiae 	& 5361 							& 63			\\
		\hline		
		S.aureus 		& 2810 							& 160			\\
		\hline
		DREAM5-Insilico & 1643 							& 850			\\
		\hline
	\end{tabular}
\end{table}

\subsection{Performance Comparison}
\label{sec:exp:others} 

\subsubsection*{Comparing Serial, Multicore, and GPU:} 

The speedup gained by multicore and GPU implementations over serial implementations are compared in Table~\ref{tab:overall}. In specific, the last column in Table~\ref{tab:overall} compares three average speedup ratios. The details are discussed below.

The first two rows in Table~\ref{tab:overall} report runtime of Stable and Parallel-PC. It is noteworthy to mention that Parallel-PC has two modes. In every benchmark, both modes are executed and the smaller runtime is reported. 
Runtime of Stable ranges from $11$ minutes in \texttt{NCI-60} to about $3$ days in \texttt{DREAM5-Insilico}. Parallel-PC takes about $11$ hours in \texttt{DREAM5-Insilico}, which is $6.7$~X faster than Stable. On average, Parallel-PC on eight cores is about $5.6$~X faster than Stable.

The third row in Table~\ref{tab:overall} reports runtime of Stable.fast on a single core. The runtime ranges from $74$ seconds in \texttt{NCI-60} to more than $11$ hours in \texttt{DREAM5-Insilico}. 
The multi-threaded mode in Stable.fast is not yet optimized at the time of this writing. With full optimizations, the multi-threaded mode may reach linear speedup gain on multicore systems. In other words, the speedup gain on eight cores compared to serial execution may reach up to $8$~X.

The fourth and fifth rows in Table~\ref{tab:overall} report runtime of cuPC-E and cuPC-S, respectively. Note that the time it takes to transfer data to and from GPU is counted as well.
Runtime of cuPC-E ranges from $440$ milliseconds to about $48$ seconds. On average, the speedup ratio of cuPC-E over the serial execution in \texttt{C} language, i.e., Stable.fast, is $525$~X. Runtime of cuPC-S ranges from $390$ milliseconds to $4.76$ seconds. On average, the speedup ratio of cuPC-S over serial execution is $1296$~X.

\begin{figure*}[tp]
	\centering
	\includegraphics[width = .7\textwidth]{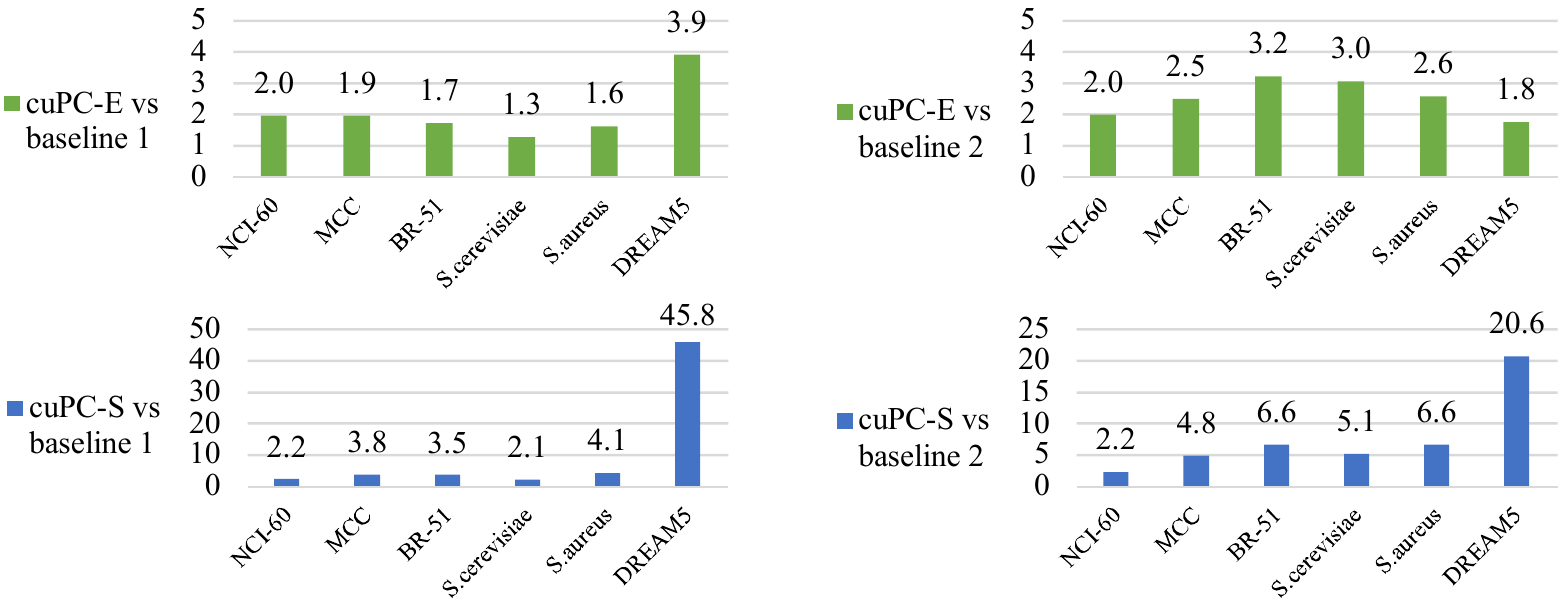}
	\vskip -3mm	
	\caption{Comparing the performance of cuPC-E and cuPC-S with two baseline GPU-parallel algorithms. Every bar represents a ratio between two runtime values. For instance, the bottom-right bar means cuPC-S is $20.6$~X faster than baseline algorithm 2 in DREAM5-Insilico dataset. 
	}
	\label{fig:exp:baseline}	
\end{figure*} 

\begin{figure*}[tp]
	\centering
	\includegraphics[width = \textwidth]{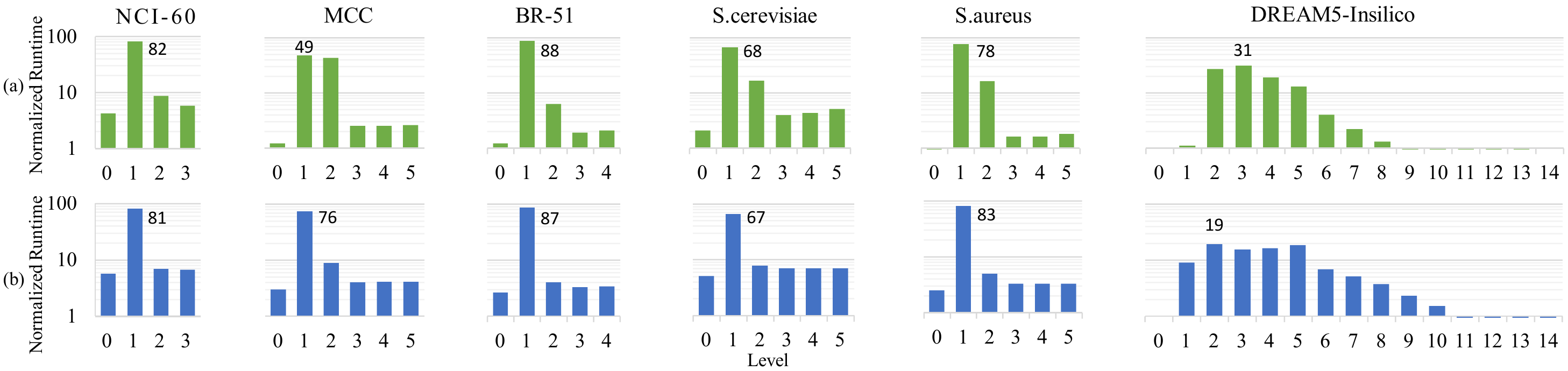}
	\vskip -3mm
	\caption{Distribution of the runtime (in percent) for a) cuPC-E and b) cuPC-S, in different levels. The values are normalized to the total runtime in every benchmark.}
	\label{fig:exp:levels}	
\end{figure*}

\subsubsection*{Comparing cuPC with Baseline Methods:} 
\mRefFig{fig:exp:baseline} compares cuPC with the following two baseline GPU-parallel algorithms. 
The first algorithm is formed by basically porting parallel-PC \cite{le2015fast} from its original multi-threaded CPU implementation to GPU. 
In specific, in every level $\mL$, all rows $i$ of the adjacency matrix are processed in parallel in separate blocks. In block $i$, all edges $(V_i,V_j)$ are processed in parallel. All the CI tests for an edge $(V_i,V_j)$ are performed sequentially in the corresponding thread. 
We also apply the same ideas in cuPC, namely, using the same compacted form of the adjacency matrix, using the same shared memory allocations, and using the same early termination strategies. 

The second baseline algorithm is formed as the following. In every level $\mL$, all elements $ij$ of the adjacency matrix, i.e., all edges $(V_i,V_j)$, are processed in parallel in separate blocks. In block $ij$, all CI tests of edge $(V_i,V_j)$ are processed in parallel. Again, the same compact, shared memory, and early termination strategies are also applied.

As illustrated in \mRefFig{fig:exp:baseline}, cuPC-E is $1.3$~X to $3.9$~X faster than baseline algorithm 1, and $1.8$~X to $3.2$~X faster than baseline algorithm 2. This shows that cuPC-E judiciously strikes a balance between the available degrees of parallelism and thus achieves higher performance compared to both of the baseline methods. 
cuPC-S is faster than cuPC-E. For instance in \texttt{DREAM5-Insilico}, which is the most challenging dataset, cuPC-S is $45.8$~X and $20.6$~X faster than the two baseline methods.

\subsubsection*{Comparing Different Levels:} 
\mRefFig{fig:exp:levels} shows distribution of the runtime values in different levels in cuPC-E and cuPC-S. Note that the reported runtime of every level includes all the corresponding overheads such as forming $A'_G$. 
In the first five benchmarks, level $1$ takes between $49\%$ to $83\%$ of the total runtime. However, in the last benchmark, level $1$ takes less than $10\%$, but levels $2$ to $5$ take $90\%$ and $70\%$ of the total runtime in cuPC-E and cuPC-S, respectively. 
This figure shows that the computations in all levels contribute to the total runtime.

\begin{figure*}[tp]
	\centering
	\includegraphics[width = \textwidth]{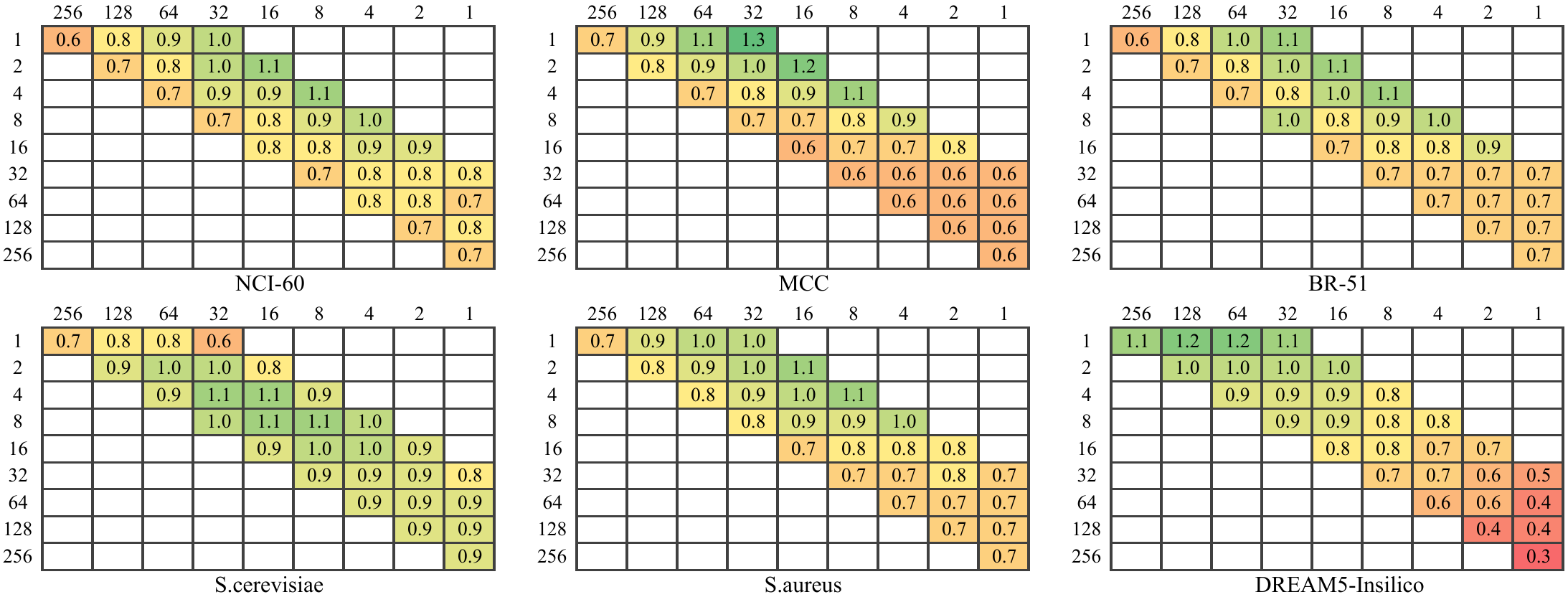}
	\vskip -3mm
	\caption{Comparing different configurations of cuPC-E with the selected configuration ($\mEB=2$ and $\mET=32$). The Y axis is $\mEB$ and the X axis is $\mET$. Green color means higher speed and red color means lower speed.}
	\label{fig:exp:he}	
\end{figure*}

\begin{figure*}[tp]
	\centering
	\vskip -3mm
	\includegraphics[width = \textwidth]{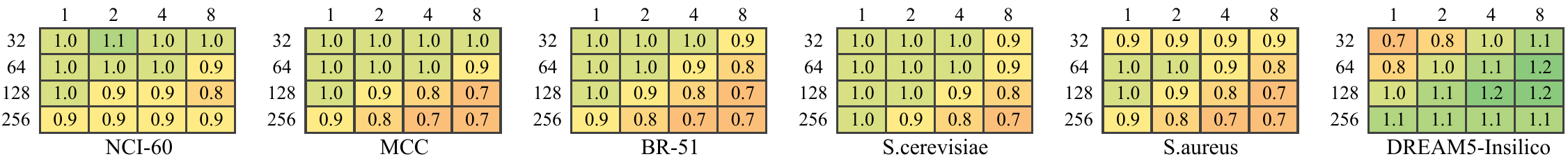}
	\vskip -3mm
	\caption{Comparing different configurations of cuPC-S with the selected configuration ($\mST=64$ and $\mSB=2$). The Y axis is $\mST$ and the X axis is $\mSB$. Green color means higher speed and red color means lower speed.}
	\vskip 0.5mm
	\label{fig:exp:hs}	
\end{figure*}

\subsection{Configuration Parameters}
\label{sec:exp:config}

cuPC-E and cuPC-S have configuration parameters which can be adjusted to improve the performance. The above results are based on executing cuPC-E with $\mEB=2$ and $\mET=32$, and cuPC-S with $\mST=64$ and $\mSB=2$. We denote these selected configurations as cuPC-E-2-32 and cuPC-S-64-2. The effect of different configurations on the performance of cuPC-E and cuPC-S is evaluated in this section.

\vskip 1mm
\textit{cuPC-E:} $30$ different configurations are experimented for cuPC-E. In specific, $\mET$ and $\mEB$ are selected from the set $\{1,2,4,\ldots,128,256\}$ such that $32 \leq \mET \times \mEB \leq 256$. 
This bounds the number of threads in every block from $32$ to $256$. Note that the number of blocks in cuPC-E is equal to $n \times \nicefrac{n'}{\mEB}$ and the number of threads in every block is equal to $\mET \times \mEB$.

The heat maps in \mRefFig{fig:exp:he} show the performance improvement or degradation of cuPC-E with different configurations compared to the selected configuration. 
The heat maps show a variation between $0.3$~X to $1.3$~X. This is mainly due to the underlying graph structure in the benchmark datasets. 
In particular, in denser graphs, the number of adjacent nodes is larger, and therefore, the number of CI tests required for every edge grows. As a result, in every row in the heat maps, configurations with larger $\mET$ show higher performance because more CI tests are executed in parallel. Note that the number of threads for the CI tests of an edge in cuPC-E is equal to $\mET$. 
For instance, in \texttt{DREAM5-Insilico}, cuPC-E-4-64 shows $10\%$ higher performance compared to cuPC-E-4-8 because $64$ threads are assigned to the CI tests of an edge instead of $8$.  
Note that there is a limit to this gain. In \texttt{DREAM5-Insilico}, the configuration 1-256 shows $10\%$ lower performance compared to 1-128 because large number of parallel threads result in too many unnecessary CI tests.  

As opposed to dense graphs, in sparse graphs higher performance is achieved in configurations with smaller $\mET$ in every row in the heat maps. For instance, in \texttt{NCI-60}, cuPC-E-2-128 shows $40\%$ lower performance compared to cuPC-E-2-16.

\vskip 1mm
\textit{cuPC-S:} $16$ different configurations are experimented for cuPC-S. In specific, $\mST \in \{ 32,64,128,256 \}$ and $\mSB \in \{ 1,2,4,8 \}$. Note that the number of blocks in cuPC-S is equal to $n \times \mSB$ and the number of threads in every block is equal to $\mST$. 
\mRefFig{fig:exp:hs} shows the performance improvement or degradation of cuPC-S with different configurations compared to the selected configuration. 

The heat maps in \mRefFig{fig:exp:hs} show a variation between $0.7$~X to $1.2$~X. Hence, compared to cuPC-E, cuPC-S shows less variation to the configuration parameters. 
This is mainly because in cuPC-S, threads are assigned to the conditional sets $S$ instead of the edges. In other words, the number of adjacent nodes of $V_i$, i.e., $n'_i$, and hence, $n'$ varies in dense or sparse graphs. This causes imbalance workloads in different blocks in cuPC-E. However, in cuPC-S, since $\binom{n'_i}{\mL}$ is normally much larger than $n'_i$, blocks are fully loaded and their workloads are more balanced.

\subsection{Global Sharing vs Local Sharing in cuPC-S}
\label{sec:exp:GlobalVsSemi}

As mentioned at the end of \mRefSec{sec:alg}, conditional sets $S$ can be shared either locally or globally in cuPC-S in order to save redundant computations and increase the overall speed. 
We employ a local sharing strategy in which only the CI tests from one row in $A'_G$ share a set $S$.  Global sharing among all CI tests from the entire graph is time consuming because it requires searching the entire graph to find all such CI tests. 
The amount of extra savings yielded by global sharing is not large enough to justify the additional cost of global search. In this section, we experimentally show the above point.

\begin{figure}[tp]
	\centering
	\includegraphics[width=0.95\columnwidth]{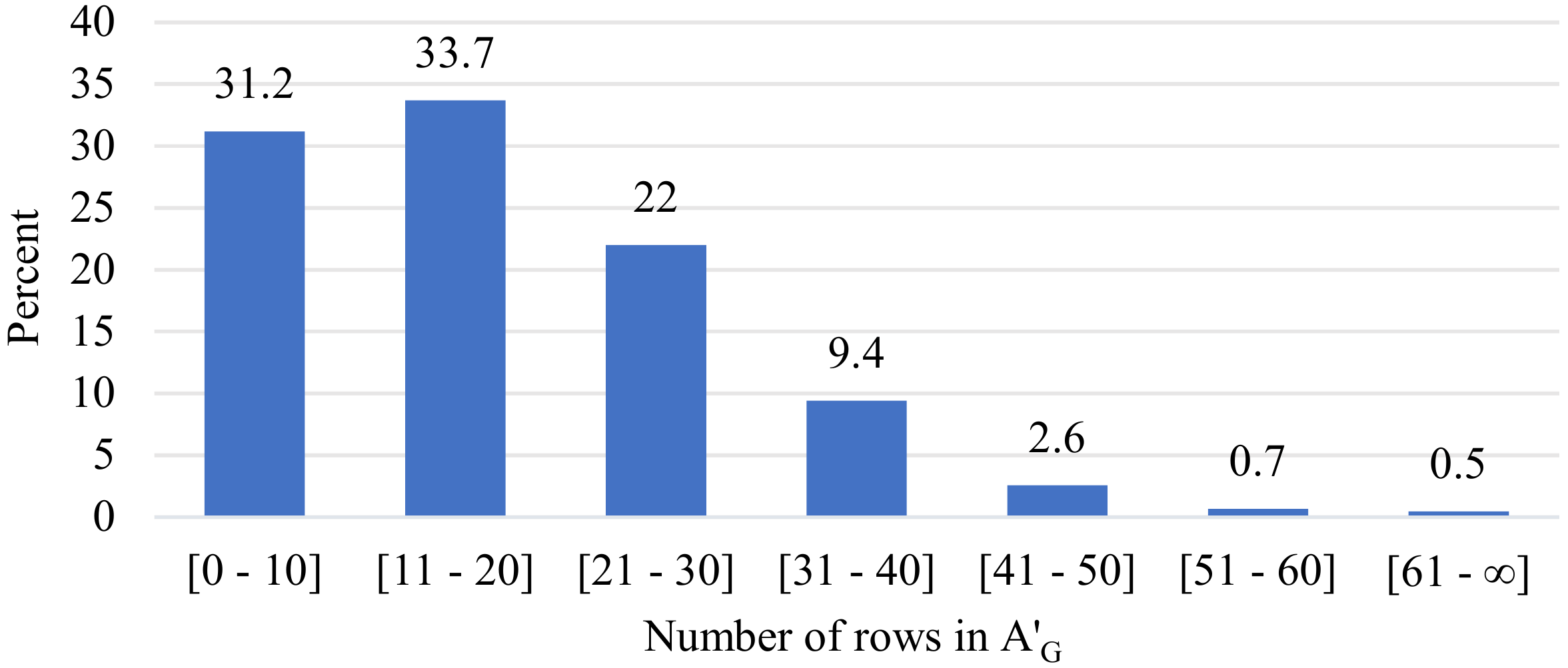}
	\vskip -3mm
	\caption{The percentage of redundant conditional sets $S$ in the entire graph in level $2$ of DREAM5-Insilico dataset. See \mRefSec{sec:exp:GlobalVsSemi} for further details.	
	}
	\label{fig:SemiVsFullReusing}	
\end{figure}

\mRefFig{fig:SemiVsFullReusing} shows a histogram. The value of each bin $[b_i,b_{i+1})$ is equal to  the number of conditional sets $S$ that appear in CI tests from at least $b_i$ to at most $b_{i+1}-1$ rows of $A'_G$ in level $2$ in \texttt{DREAM5-Insilico} dataset. 
The figure shows that about $95\%$ of the redundant conditional sets $S$ appear in at most $40$ rows of $A'_G$. This is much smaller than the total number of rows in this dataset, i.e., $n=1643$. 
Hence, the cost of global search is not justified.

\begin{figure}[tp]
	\centering
	\includegraphics[width=0.9\columnwidth]{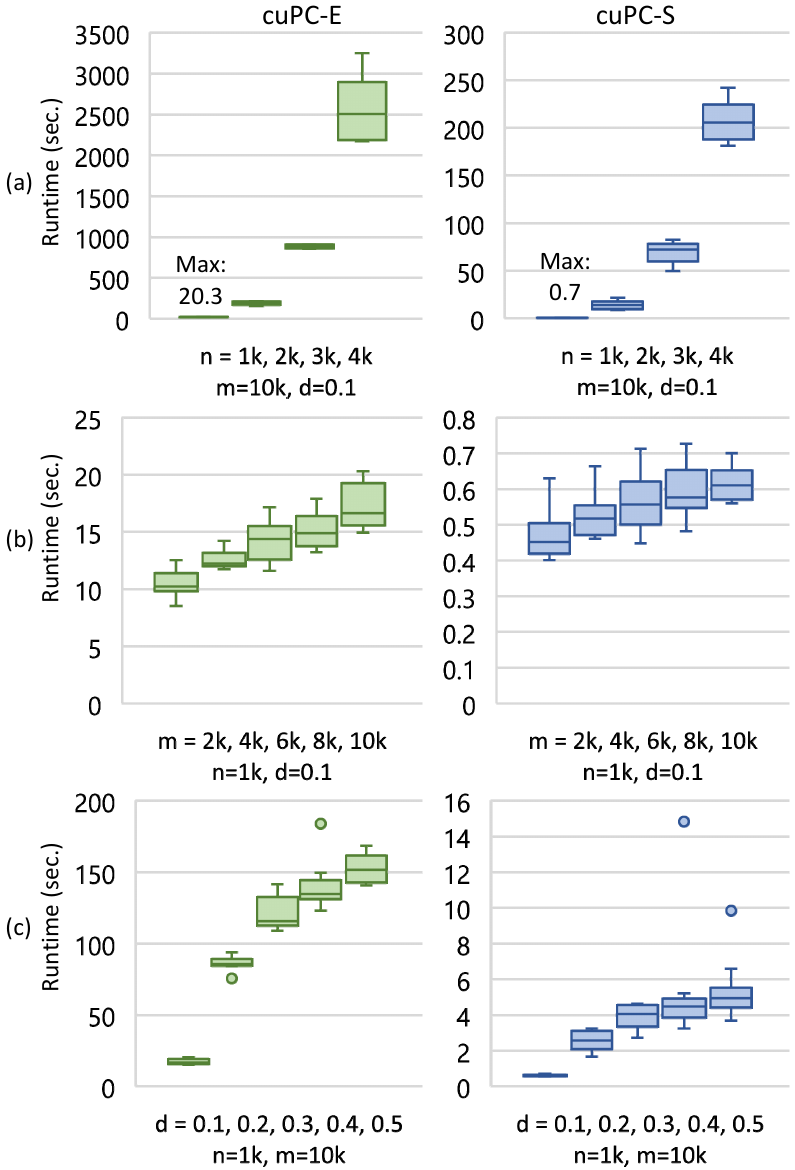}
	\vskip -2mm
	\caption{
		Runtime of cuPC-E and cuPC-S with a) different number of variables, b) different sample sizes, and c) different graph densities. 
		Every box-and-whisker plot shows quartiles 1, 2 (median), and 3, plus the lowest point still within $1.5$ IQR of the lower quartile, and the highest point still within $1.5$ IQR of the upper quartile. The outliers are shown as small circles.
	}
	\label{fig:exp:scale}	
\end{figure}

\subsection{Scalability}

Scalability of the proposed parallel algorithms are evaluated in this section. In specific, performance of cuPC-E and cuPC-S are experimented for different number of variables ($n$), different number of samples ($m$), and different graph densities ($d$).

To evaluate the impact of scaling the number of variables, we consider $n=1000$, $2000$, $3000$ and $4000$. In every case, ten graphs are generated by randomly drawing an edge between any pairs of variables with probability $d=0.1$. 
In particular, we first generate a random adjacency matrix $A_G$ with independent realizations of Bernoulli random variable with parameter $d$ in the lower triangle of the matrix and zeros in the remaining entries. Next, we replace the ones in $A_G$ by independent realizations of a uniform random variable in the range $[0.1,1]$. A non-zero entry $A_G[i,j]$ shows that there is a direct causal effect from $V_j$ to $V_i$. 
Next, from $i=0$ to $i=n-1$, i.e., from top to bottom, the samples are generated as $V_i = N_i + \sum_{j=0}^{i-1} A_G[i,j] V_j$, where the random variables $N_i$'s have normal distribution and are mutually independent. The sample size for every random variable is set to $m=10000$.

Next, cuPC-E and cuPC-S are executed and the runtimes are measured in every case. The results are shown in \mRefFig{fig:exp:scale}(a). Runtime increases with $n$, but cuPC-S always has higher performance compared to cuPC-E.  
We also executed the C implementation of PC-stable on the same datasets. However, even in the smaller graphs ($n=1000$), PC-stable could not produce results after $48$ hours, and thus, we aborted the job. Hence, cuPC-E is at least $48 \times 3600 ~/~ 20.3 \text{~sec.} \simeq 8500$ X faster than PC-stable in this case.

Next, the impact of scaling the sample size is experimented. We consider $m=2000$, $4000$, $6000$, $8000$, and $10000$. Here, $n=1000$ and $d=0.1$. In every case, ten random graphs are generated as discussed above and runtimes are measured. The results are shown in \mRefFig{fig:exp:scale}(b). 
The runtime increases linearly with the sample size. Increasing the sample size, improves the accuracy of the CI tests. This decreases the number of edges that are removed in level $\mL$, which in turn, increases the number of CI tests required to be performed in level $\mL+1$.

Finally, the impact of scaling the graph density is experimented. We consider $d=0.1$, $0.2$, $0.3$, $0.4$, and $0.5$. Here, $n=1000$ and $m=10000$. The results are shown in \mRefFig{fig:exp:scale}(c). 
Increasing $d$ means the graph is more dense, the number of remaining edges are increased, and hence, the runtime should increase.  
Runtime of cuPC-E and cuPC-S increase almost linearly from density $0.2$ to $0.5$. However, at density $0.1$, the runtime is much smaller. This is because the runtime changes by optimizing the configuration parameters in every case, while we employ the same configuration across all values of $d$. Therefore, in some cases, e.g., in $d=0.1$, the selected configuration is a better fit and the algorithm runs faster.

\section{Conclusion}
\label{sec:conc}

In empirical sciences, it is often vital to recover the underlying causal relationships among variables in real-world high-dimensional datasets. In this paper, we proposed a GPU-based parallel algorithm for PC-stable with two variants, i.e., cuPC-E and cuPC-S, to learn causal structures from observational data. Experiments showed the scalability of our prospered algorithms with respect to the number of variables, the number of samples, and different graph densities. Note that the proposed solution also helps to accelerate some other causal structure learning algorithms such as CCD, FCI, and RFCI, because they use PC algorithm as a subroutine. 





%

\ifCLASSOPTIONcaptionsoff
  \newpage
\fi



%

%



\newpage

\begin{IEEEbiography}[{\includegraphics[width=1in,height=1.25in,clip,keepaspectratio]{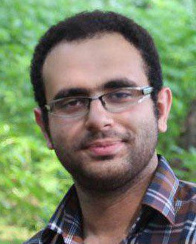}}] {Behrooz Zarebavani} received the B.Sc. degree in electrical engineering from Amirkabir University of Technology, Tehran, Iran, in 2017. He is currently working towards the M.Sc. degree in electrical engineering at Sharif University of Technology, Tehran, Iran. His research interests include parallel processing, machine learning, and casual inference. 
\end{IEEEbiography} 
\begin{IEEEbiography}[{\includegraphics[width=1in,height=1.25in,clip,keepaspectratio]{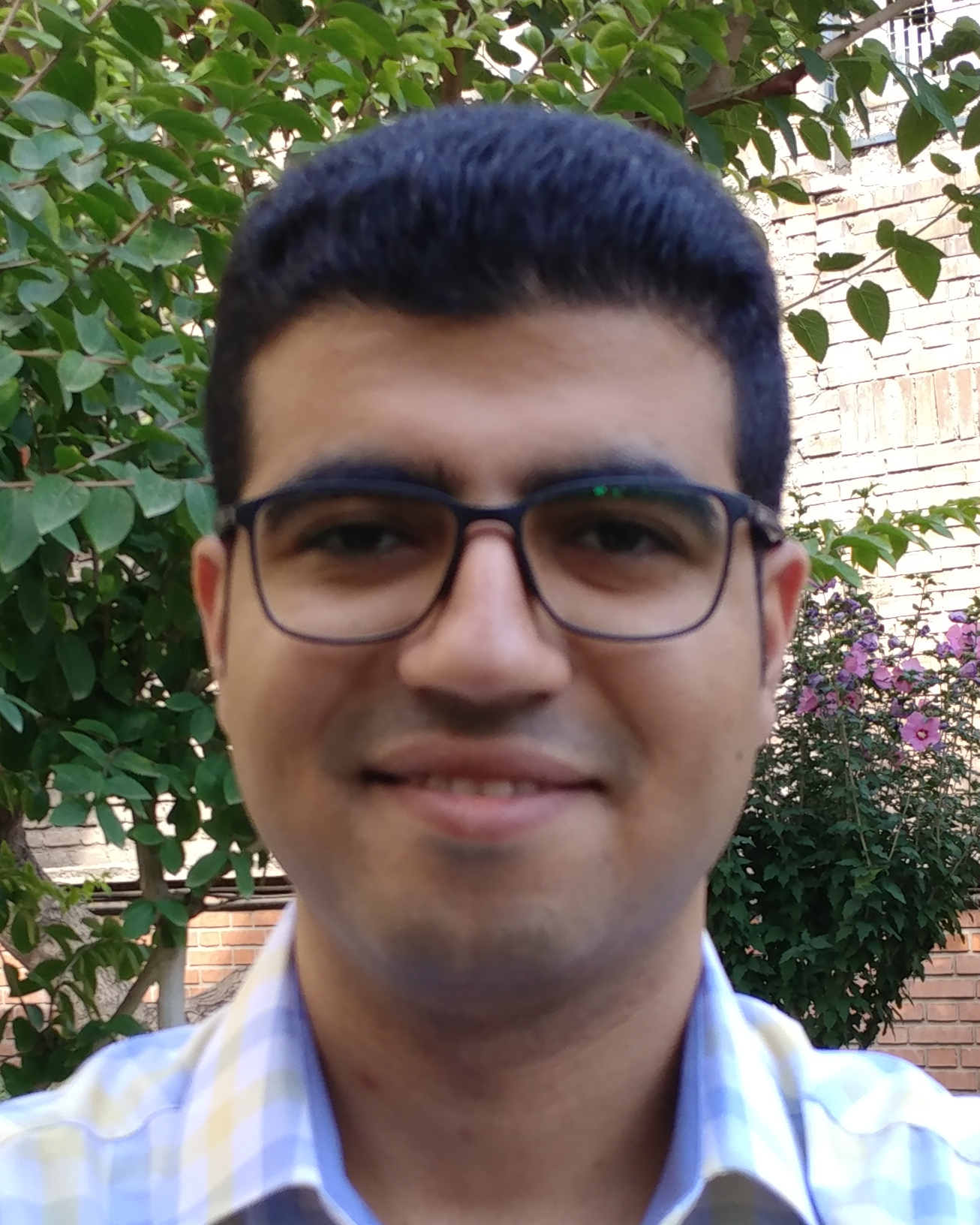}}] {Foad Jafarinejad} received the B.Sc. degree in electrical engineering from Sharif University of Technology, Tehran, Iran, in 2019. His research interests include machine learning, graphical model learning, and parallel computing.
\end{IEEEbiography} 
\begin{IEEEbiography}[{\includegraphics[width=1in,height=1.25in,clip,keepaspectratio]{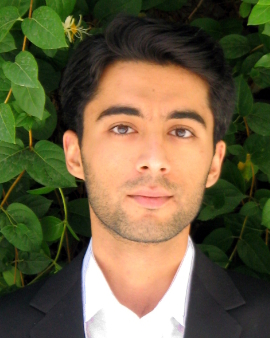}}] {Matin Hashemi} received the B.Sc. degree in electrical engineering from Sharif University of Technology, Tehran, Iran, in 2005, and the M.Sc. and Ph.D. degrees in computer engineering from University of California, Davis, in 2008 and 2011, respectively. He is currently an assistant professor of electrical engineering at Sharif University. His research interests include algorithm design and hardware acceleration for machine learning, signal processing, and big data applications.
\end{IEEEbiography} 
\begin{IEEEbiography}[{\includegraphics[width=1in,height=1.25in,clip,keepaspectratio]{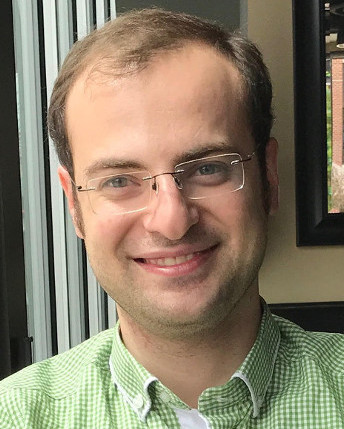}}] {Saber Salehkaleybar} received the B.Sc., M.Sc. and Ph.D. degrees in electrical engineering from Sharif University of Technology, Tehran, Iran, in 2009, 2011, and 2015, respectively. He is currently an assistant professor of electrical engineering at Sharif University of Technology. His research interests include distributed systems, machine learning, and causal inference.
\end{IEEEbiography}





\end{document}